\documentclass[aps,pre,twocolumn,showpacs,amssymb,floatfix,superscriptaddress,letter]{revtex4-1}
\input epsf

\begin{document}
\title{A stroll in the energy landscape}

\author{Antonio Scala}
\author{Luca Angelani}
\affiliation{INFM and Dipartimento di Fisica, 
Universit\'a di Roma {\rm La Sapienza}, 
I-00185, Roma, Italy}

\author{Roberto Di~Leonardo}
\affiliation{INFM and Dipartimento di Fisica,
Universit\'a di L'Aquila, 
I-67100, L'Aquila, Italy}

\author{Giancarlo Ruocco}
\author{Francesco Sciortino}
\affiliation{INFM and Dipartimento di Fisica, 
Universit\'a di Roma {\rm La Sapienza}, 
I-00185, Roma, Italy}

\begin{abstract}
We review recent results on the potential energy landscape (PES) of
model liquids. The role of saddle-points in the PES in connecting
dynamics to statics is investigated, confirming that a change between
minima-dominated and saddle-dominated regions of the PES explored in
equilibrium happens around the Mode Coupling Temperature. The
structure of the low-energy saddles in the basins is found to be
simple and hierarchically organized; the presence of saddles nearby in
energy to the local minima indicates that, at non-cryogenic
temperatures, entropic bottlenecks limit the dynamics.
\end{abstract}

\maketitle

The study of the properties of the free energy landscape and/or of the
potential energy surface (PES) and their connection with the slow
dynamics in undercooled liquids and disordered systems is a topic of
current research~\cite{tutti}. The original idea of
Goldstein~\cite{goldstein} of considering the atomic dynamics of
undercooled liquids as the incoherent superposition of a vibrational
motion confined in the basin of attraction of local minima of the PES,
interspersed by jumps among the basin of different minima, was later
developed in a formal theory by Stillinger and
Weber~\cite{stillinger}.  These authors define the basin of attraction
of a given local minimum of the PES as the set of all that points in
configuration space who are linked to the minimum through a steepest
descent path. It is obvious that -~according to this definition~- the
basins of attraction generate a complete partition of the
configurational space. Consequently, Stillinger and Weber write the
partition function as
\begin{equation}
Z=\sum_{\{B\}} Z_{B}
\end{equation}
\noindent where $\{B\}$ indicates the set of all the basins and
$Z_{B}$ is the partition function restricted to the basin $B$.  The
points belonging to the borders $\{\partial B\}$ of the basins are
safely excluded from the calculation of the partition function, as
they form a set of zero measure.

In this way Stillinger and Weber succeeded to write the free energy of
a system with a continuous PES as the sum of the free energy of a
typical basin $F_{basin}(T,\rho)$ and a term accounting for the
multiplicity of basins accessible by the system $\Omega(T,\rho) =
\exp(S_{conf}(T,\rho)/k_B)$. Excluding the local minima corresponding
to crystalline structures, they arrive to an expression for the total
free energy of the liquid phase:
\begin{equation}
F_{liquid}=F_{basin}(T,\rho)-T S_{conf}(T,\rho)
\end{equation}
\noindent that can be used to predict, for example, the melting
temperature $T_m$ as that temperature where $F_{liquid}$ becomes equal
to the free energy of the crystal $F_{crystal}$.

\begin{figure}
\hbox to\hsize{\epsfxsize=0.8\hsize\hfil
\epsfbox{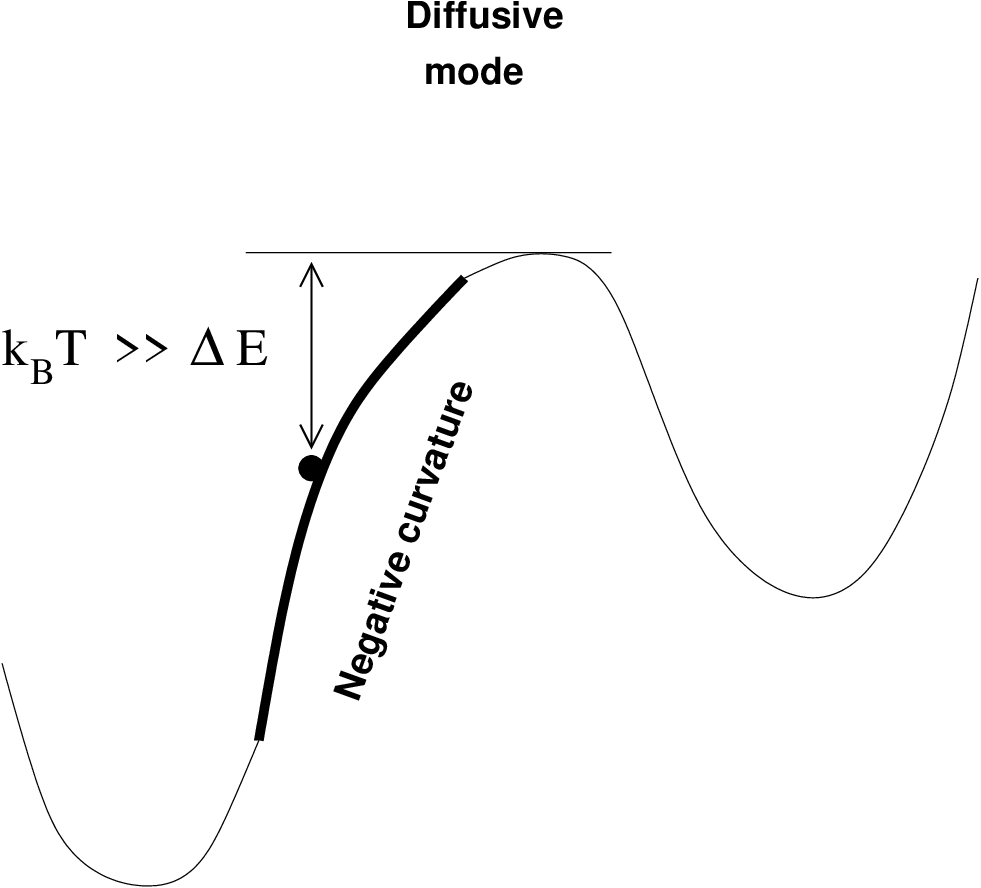}\hfil}
\hbox to\hsize{\epsfxsize=0.6\hsize\hfil
\epsfbox{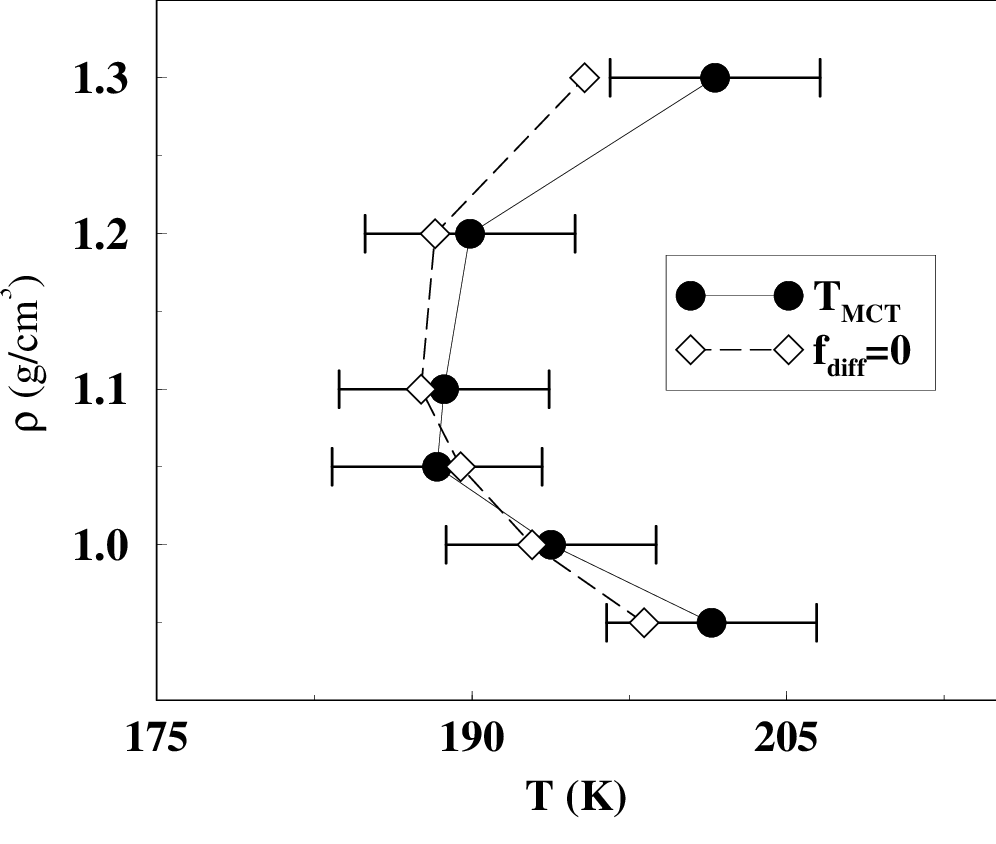}\hfil}
\caption{
a) Sketch of the potential energy surface (PES) along a direction with
negative curvature that could lead to inter-basin diffusion. The
marked point represents the position of the system; for $T>T_{MCT}$
the apparent energy barrier between this point and the top of the
$PES$ along the direction considered is less than $\sim 10^{-2}$
$k_B T$ for the models studied. Although the section of the PES in
this example has two minima along the direction of the eigenmode
associated with this negatively curved direction, it can be that these
apparent minima lead to the same basin~~~~ b)~The fraction
$<f_{diff}>$ of diffusive modes in the SPC/E model for water goes to
zero at the mode-coupling transition temperature $T_{MCT}$ (from
ref.~\protect\cite{INMwater}).  }
\label{figINM}
\end{figure}

According to the Stillinger and Weber idea, the liquid thermodynamics
can be derived from the knowledge of the {\it basins} of the PES. On
the contrary, the proposal of Keyes to use the instantaneous normal
mode (INM) theory to predict not only the very short time dynamics but
also the long time diffusion processes~\cite{keyes} gives relevance to
the process of crossing the borders of the basins and, therefore, to
the structure of these borders. In the INM approach, the local
curvature of the PES is calculated along equilibrated molecular
dynamics (MD) trajectories. The idea is to reconstruct the global
shape of the landscape explored from local information. For example,
if the system is near enough to a local minimum, the local curvatures
of the PES will be all positive and similar in value to the curvatures
of the nearby minimum. On the other hand, points near the borders
between basins will have negative curvatures along some directions. In
fact, the potential energy has a maximum along any direction crossing
a border (Fig.~\ref{figINM}a).  Although the exact functional form is
system dependent~\cite{funcformINM}, the INM approach suggests that
the diffusion coefficient $D$ is a monotonic function of the fraction
of unstable modes $<f_u>$=$<N_u>/N$, where $<N_u>$ is
the number of negative eigenvalues of the Hessian matrix evaluated
along the actual MD trajectory.

A more quantitative relation between the number of negative curvature
of the PES and the diffusion coefficient stems from the observation
that not all the directions with negative curvatures lead to a change
of basins: even in the crystalline state, a non-zero $<$$f_u$$>$ is
found due to the anharmonicities of the potential. Depending on the
different authors and on the different model studied, different
methods to discriminate the fraction $<$$f_{diff}$$>$ of truly
diffusive directions from the whole set of negative curvatures have
been proposed. A non zero value of $<$$f_{diff}$$>$ indicates that the
trajectory followed by the system is confined to regions of the
configuration space {\it near} the borders of the
basins~\cite{nota_A}.

The study of the diffusion coefficient in undercooled liquids
naturally leads to a comparison with the Mode-Coupling Theory
(MCT)~\cite{goetze1}. Indeed, the ideal MCT predicts the existence of
a finite dynamical critical temperatures $T_{MCT}$ at which $D=0$.
The ideal mode-coupling transition can be derived under the restricted
assumptions of the gaussianity of the
fluctuations~\cite{mctGAUSS}. The INM approach reveals that at
$T_{MCT}$ there is a change in the temperature behaviour of the
quantity $<$$f_{diff}$$>$: it becomes numerically zero
(Fig.~\ref{figINM}b). Therefore one can conclude that at $T_{MCT}$ the
system ceases to be near the borders of the basins and starts to spend
most of the time close to the minima.

On a general ground, one has to understand why the system is not
already close to the minima at undercooled temperatures
$T_{MCT}<T<T_m$, or, in other words, why the slowing down of the
diffusion processes takes place even if the system spend most of the
time close to the basin borders. The analysis of Kurchan and
Laloux~\cite{kurchanlaloux} on the properties of borders in
high-dimensional configuration spaces indicates a simple reason for
this behaviour. In a $N$-dimensional space, (i.e., the configuration
space of a system with $N$ degrees of freedom) a basin $B$ has a
volume at least of the order $l^N$, where $l$ is a finite length and
is the typical distance among configurations in the same basin. We can
imagine then to give a fictious volume to the border $\partial B$
considering ``belonging to $\partial B$'' all the points in a shell of
width $\epsilon$ and volume $\sim \epsilon N l^{N-1}$ near $\partial
B$. The fraction of border points is therefore $\epsilon N / l$
indicating that in the thermodynamic limit $N \rightarrow \infty$ most
of the volume of the basins is located in a shell of width $\epsilon
\sim 1/N $ near the borders. The numerically observed decrease of
$<$$f_{diff}$$>$ with the temperature gives us further indications on
the structure of the borders. At high temperatures, the system
populates regions of the borders that are in common to many basins,
while near $T_{MCT}$ the system is near regions of the borders in
common to few basins; therefore the regions of the borders in common
to few basins have a smaller volume compared to regions in commons to
many basins.  This seems to indicate a hierarchical organization of
the configuration space~\cite{kurchanlaloux}. Indeed, let's assume
that we are able to study the potential energy $U$ restricted to the
borders $\partial B$ of the basins. This restricted PES is an
$(N-1)$-dimensional manifold where all the stationary
points~\cite{statpoint} have their order decreased by one: so, the
saddles of order one become the local minima of the restricted
$(N-1)$-dimensional PES and partition this manifold in basins. Again,
if the system is at high temperatures, the trajectories will be
localized near the $(N-2)$-dimensional borders of the
$(N-1)$-dimensional basins of attractions; so, one can iterate the
procedure until one consider the relevant $(N-i)$-dimensional manifold
in the nearby of which equilibrium trajectories are ``almost always''
within some small distance (with the order $i$ less than or equal to
$N<f_{diff}>$).

Following the previous rationale, in a rough way, one could write the
partition function of the system as
\begin{equation}
\label{partinsaddles}
Z=\sum_{i}Z_{M_i}
\end{equation}
\noindent where $M_i$ is the set of points in common to $i$
basins and $Z_{M_i}$ is the partition function restricted to the
$(N-i)$ dimensional manifold $M_i$. The local minima of the potential
energy $U$ restricted to $M_i$ are the saddles of order $i$ in the
unrestricted configuration space. Although from the strict point of
measure the manifolds $M_i$ have zero measure, we know from the INM
studies that at temperatures $T>T_{MCT}$ the system is in the nearby
of some manifold $M_{i(T)}$ with $i(T)>0$. In order to give an
operational definition of such partition of the space in basins of
attractions of the saddles, one has to give a finite volume to the
borders. To this aim, two recent studies~\cite{angesaddle,cavagnone}
have used the pseudo-potential introduced by
Stillinger~\cite{stillgradU}:
\begin{equation}
\Phi= \vert \vec \nabla U \vert^2.
\end{equation}
\noindent

It is worth to note that the stationary point of any order of the PES
are absolute minima of such potential (in particular they are
characterized by $\Phi=0$), but there exist minima of $\Phi$ -~those
with $\Phi\neq 0$~- that are not stationary point of $U$.  This
implies that the use of $\Phi$ is not rigorously correct if the goal
is to obtain a partition of the configuration space in basins of
attractions of saddles of different orders.

The choice of $\Phi$ provides anyhow a partitioning of the configuration
space in basins of attraction of its local minima. As shown recently
in~\cite{wales} and confirmed by a more detailed analysis by
us~\cite{newangelani}, the majority of local minima of
$\Phi$ are not true saddles but present a 2\% fraction of directions
which are indeed inflection points.  Despite this, the T-dependence of
the number and energy location of these ``quasi-saddles'' carries
informations on the dynamical slowing down upon
cooling~\cite{angesaddle}. Therefore, although not all the minima of
$\Phi$ found can be classified as ``true'' saddles~\cite{wales}, we
will still assume that they are a good sampling of the PES and
therefore that the ``true'' saddles maintain the same characteristics
(i.e. curvatures, number of negative directions, energy) of the found
local mimima of $\Phi$; in this respect, we will continue using the
word "saddles" in this manuscript, keeping in mind that we refer to
such minima.

For the case of simple model liquids the order $<$$N_s$$>$ and the
energy of the saddles explored during the MD simulations have
determined~\cite{angesaddle,cavagnone}. As a result, the saddle's
order has been found to be a decreasing function of the temperature
and to become zero at the $T_{MCT}$ temperature of the system under
considerations. This result confirms the role of $T_{MCT}$ as a
crossover temperature from a dynamics near the basins's borders
($<$$N_s$$>$ $>$ 0) to a dynamics mainly localized inside the basins
($<$$N_s$$>$ $=$ 0)~\cite{Schroder}. Given this findings, the strong
non-Arrhenius increase of the relaxation times of the dynamics
approaching $T_{MCT}$ from above can be explained with the decrease of
the diffusive direction, whose number is proportional to
$<$$N_s$$>$. Indeed, although the system does not need to cross any
energy barrier to ``jump the ridge'' among basin, the exploration of
the configuration space is slower as the connectivity (i.e. the number
of adjacent basins that the system can freely reach) among basins
decreases.

A direct evidence of situations in which, although the energy of the
saddles is significatively less then the energy of the system, the
system is trapped in the basin of attraction of a local minimum, is
found in~\cite{Wevers}, where the connectivity between the crystalline
ground state and the adjacent local minima is analized for a model of
the crystalline compound $MgF_2$. Moreover, it is found
in~\cite{Wevers} that, at higher energies, saddles connecting several
minima make up a sizable fraction of the transition region.

A further confirmation of the "entropic" -~rather than "energetic"~-
origin of the slowing down of the dynamics is given by numerical aging
experiments. For Lennard-Jones liquids, it has been shown that in the
aging following a deep quench, the system follows quasi-equilibrium
states~\cite{shortaging} during which short time dynamics is
intra-basin vibration, while long time dynamics is dominated by the
slow search of deeper and deeper local minima. The work of Angelani et
al.~\cite{angeaging} shows that, after a short transient, the system
changes basins crossing saddles of order one without any energy
activated process.

The result that at $T_{MCT}$ the value of $<$$N_s$$>$ approaches zero
has lead to the interpretation of the mode coupling critical
temperature as the threshold below which the diffusive dynamics
becomes {\it activated} with potential energy barriers that the system
needs to overcome in order to cross the border between
basins~\cite{cavagnino}. We show here -~after a careful definition of
the meaning of the term "activated processes" in the framework of the
PES description of the dynamics~- that it is at least unnecessary to
invoke potential energy barrier crossing, and that there is no
qualitative change in the way the system change basins below or above
$T_{MCT}$. What is qualitatively different on crossing $T_{MCT}$ is
the vanishing probability of occupying a basin of a saddle below
$T_{MCT}$.

Let's first discuss the concept of activated processes in the
framework of landscape theory. Landscape theory considers the motion
of a representative point in an $N$-dimensional configuration space;
in the thermodynamic limit $N \rightarrow \infty$, the value of
thermodynamic quantities like the average potential energy $U$ becomes
determined apart of fluctuations of order $1/\sqrt
N$~\cite{foundation}.  Although in ergodic theory the complexity of
the landscape is not explicitly considered but as the hypotesized
source of random couplings enforcing the large-numbers law and the
ergodicity~\cite{ergodic}, the apparatus of the ergodic theory
approach to the foundation of statistichal mechanics is based on the
analysis of the motion of the system on iso-potential manifolds of the
PES. The exploration of such manifolds is purely entropic, and no
fluctuation in energy is considered. Obviously, for $N$ large enough,
one can always consider a subset of the system with a finite number
$M$ degrees of freedom. This subset can be then considered as a
canonical system of $M$ degrees of freedom with finite fluctuations of
order $1/\sqrt M$~\cite{foundation}. In general, it is possible to
select a subset of coordinates ($\omega$, the "reaction" coordinates)
relevant to the process under consideration, as for example the
coordinate describing the motion across the basin's border. In this
case the potential energy {\it of the subset along the reaction
coordinate} goes through a maximum (a saddle point in the
configurational space). The other degrees of freedom can be considered
as a "thermal bath" for the reaction coordinate.

Our aim is now to show that both approaches -~the one considering the
systems as a whole moving on a constant {\it potential} surface and
the one that focus the attention on the reaction coordinate and
consider the rest of the system as a thermal bath~- can be used to
correctly determine the "transition rates", i.~e.  that quantity that
control the inter-basins dynamics.

According to the second point of view, the expression of the free
energy $F(\omega, k_B T,...)$ is found integrating over the degrees of
freedom orthogonal to $\omega$, and the result is the usual free
energy in the canonical ensemble. This free energy defines the {\it
pure states} as local minima of the free energy (a {\it pure state} is
now a not necessarily compact set of points in the configuration
space); the {\it free energy barriers} among these {\it pure states}
signal transitions among states. Yet, once a coarse-grained
description of the system in terms of reaction coordinates $\omega$ is
given, the possibility of discriminating among energy and entropy
barriers is found. As a back-of-the-envelope calculation to illustrate
the concept, let's consider a collection of $N$ harmonic degrees of
freedom plus a double well degree of freedom with two equal minima
separated by an {\it energy} barrier of height $\Delta$. The reaction
coordinate corresponds obviously to the anharmonic degree, and the
transition rate among the minima can be calculated to be $\propto
exp({- \Delta / k_B T})$. If, on the contrary, we work in the spirit
of the second point of view, i.~e. with trajectories at constant
potential energy $U$, we can estimate the transition rate as a ratio
of volumes, and specifically as the ratio of the volume $\sim (\sqrt
U)^N$ relative to the time that the system spend in the basin of a
minimum and the volume $\sim \left( \sqrt (U-\Delta) \right)^N$
pertaining to the time spent in the order-one saddle basin. In the
thermodynamic limit, where $N \rightarrow \infty$ and $U \rightarrow N
k_B T/2$, we have that the transition probability $\left( 1 - 2\Delta
/ N k_B T \right)^{N/2}$ converges to $exp( - \Delta / k_B T)$. We can
conclude that the two points of view give rise to the same expression
for the transition probability.

As a further example we present the numerical comparison between the
"true" dynamics of a finite size systems and a fictious dynamics where
the same system is forced to evolve at constant potential energy. In
particular, we studiy the self diffusion coefficient in the so called
"Trigonometric Model" (TM), introduced by Madan and
Keyes~\cite{TriModelKeyes} as a dynamical system with a very simple
structure of the PES. In the TM each degree of freedom is independent
from the other and experiences a periodic external force, in
particular the TM potential energy has the form:
\begin{equation} V(\{x\}_{i=1..N})=\sum_{i=1}^N \Delta
[1-cos(\pi x_i/d_o)].
\label{pot}
\end{equation}
Despite its simplicity, the PES of this model contains some features
that have been retrieved in more realistic glasses. The TM shares with
the binary mixture -~and with the modified~- LJ system investigated in
Ref.~\cite{angesaddle} the existence of a regular organization of the
saddles above a given minimum (the elevation in energy of the saddle
is proportional to the saddle's order) and a regular distribution of
the minima in the configurational space (nearest neighbor minima lie
at a well defined Euclidean distance). What the TM misses is the
existence of minima with different energies, a characteristic of
realistic glasses~\cite{KobSciort}. This deficiency is not important
here, as we want only to select a toy model to compare the "canonical"
and "iso-potential" dynamics. In order to simplify the calculation, in
this example we treat the dynamical evolution of the system following
a Langevin dynamics in the high friction limit.

\begin{figure}
\hbox to\hsize{\epsfxsize=0.8\hsize\hfil
\epsfbox{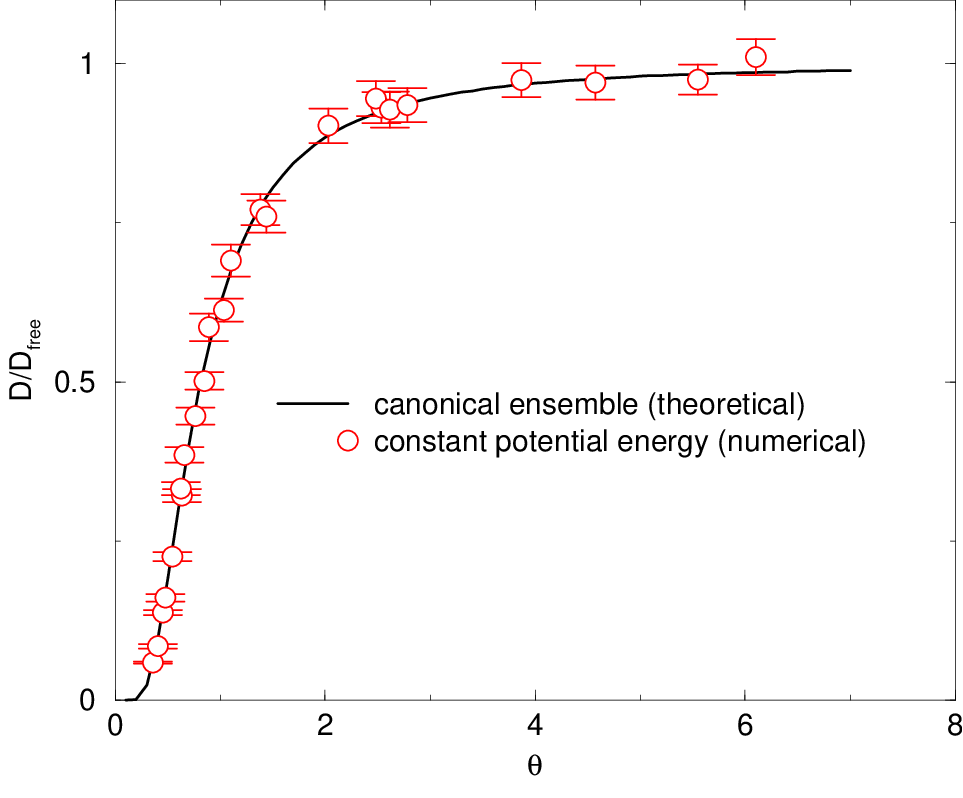}\hfil}
\caption{ 
Diffusion in the trigonometric model~\protect\cite{TriModelKeyes}.
The continuous line is the analytical solution for the dynamics in the
canonical ensemble~\protect\cite{Risken} ; circles are numerical
calculations for the constant potential energy
dynamics~(Eq.~\protect\ref{Lang2}). }
\label{figDIFFTRIG}
\end{figure}

In a first step we evaluate the time evolution of the $N$ variables
$x_i(t)$ according to the true Langevin dynamics:
\begin{equation}
{\dot x}_i(t)=-\Gamma \nabla_i V+\eta_i \label{Lang}
\end{equation}
where $\eta_i(t)$ is a random variable, $\delta$-correlated in time
and with variance proportional to the temperature,
$<\eta_i(t)\eta_j(t')>=2\delta_{ij}\delta(t-t')\Gamma k_BT$, and
$\Gamma$ fixes the time-scale of the evolution. In the case of the
stochastic motion in periodic potential (as the TM), the self
diffusion coefficient $D=lim_{t\rightarrow\infty} (N\; t)^{-1} \sum_i
<\vert x_i(t)-x_i(0)
\vert^2>$ can be calculated analytically~\cite{Risken} and turns out
to be:
\begin{equation} D=\frac{\Gamma}{\beta}
\frac{1}{Z(\beta)}\frac{1}{Z(-\beta)}
\label{DdaZ}
\end{equation}
where $Z(\beta)$ is the partition function for a single degree of
freedom. For potentials -~like that of Eq.~(\ref{pot})~- having a skew
symmetry, $V(x)=2\Delta-V(x-d_o/2)$, equation~(\ref{DdaZ}) simplifies
to:
\begin{equation}
D=\frac{\Gamma}{\beta} \frac{e^{-2\beta\Delta}}{Z^2(\beta)}.
\end{equation}
Finally, by calculating the partition function for the TM, one gets
\begin{equation}
D=\frac{\Gamma}{\beta} I_o^{-2}(\beta\Delta) \label{diff}
\end{equation}
where $I_o(x)$ is the Bessel function of order 0. The dimensionless
quantity $D/D_{free}$, where $D_{free}=\Gamma k_B T$ is the diffusion
coefficient for free brownian motion, is reported as a function of
$\theta=k_BT/\Delta$ in Fig.~\ref{figDIFFTRIG} (full line).

As second step, we calculate the diffusion coefficient for the same
potential model and the same stochastic dynamics, but with a constrain
that obliges the system to move at constant potential energy. The
motion at constant potential energy implies that $\sum_i{\dot x}_i
\nabla_i V =0$, and this condition is automatically satisfied by
modifying the equation of motion~(\ref{Lang}) as:
\begin{equation}
{\dot x}_i(t)=\eta_i-\frac{\sum_j (\eta_j \nabla_j V)}{\sum_j
(\nabla_j V)^2} \; \nabla_i V 
\label{Lang2}
\end{equation}
The evolution of $N$=1000 variables obeying Eq.~(\ref{Lang2}) has been
solved numerically at different temperatures. The diffusion
coefficients calculated in these runs (for each $T$ we made 500
replica each one 500$^.$000 time steps long) are reported in
Fig.~\ref{figDIFFTRIG} (open dots). The comparison between the values
of $D$ evaluated along canonical trajectories (full lines) and along
iso-potential trajectories (open circles) clearly indicates that,
already for not-too-large system sizes ($N$=1000), the "true" dynamics
take place along constant potential energy manifolds, and that the
potential energy fluctuations, that are present in the canonical
calculation, are not relevant for describing the diffusion process.

As we have shown, and substantiated with the previous examples, in the
thermodynamic limit the dynamics of the system as a whole takes place
along constant potential energy manifolds, the fluctuation becoming
irrelevant. In this framework, i.~e.  discussing the undercooled
dynamics as motion of the representative point in the configuration
space, the notion of activated process becomes irrelevant as
well. Either at high and low temperature, above or below $T_{MCT}$,
the representative point in the configuration space never crosses
potential energy barriers. The slowing down of the dynamics with
decreasing $T$ is a consequence of a topological properties of the
PES, i.~e. the decreasing number of the accessible iso-potential paths
between minima with decreasing potential energy. Some of these open
paths still exist even on the typical constant potential energy
manifolds visited at temperature well below $T_{MCT}$, as shown, for
example, in Ref.~\cite{angesaddle}. One can conclude that the
dynamical slowing down is fully entropy controlled and no {\it
qualitative} differences in the way the system changes basins are
observed crossing $T_{MCT}$.  On the contrary, a {\it qualitative}
change in the dynamic is expected to take place at even lower $T$ (at
the Kautzmann temperature $T_K$), where the number of accessible
basins becomes non extensive. In this respect, whether or not $T_K$ is
different from zero is still matter of debate.


\newpage

\newpage

\begin{figure}
\hbox to\hsize{\epsfxsize=0.8\hsize\hfil
\epsfbox{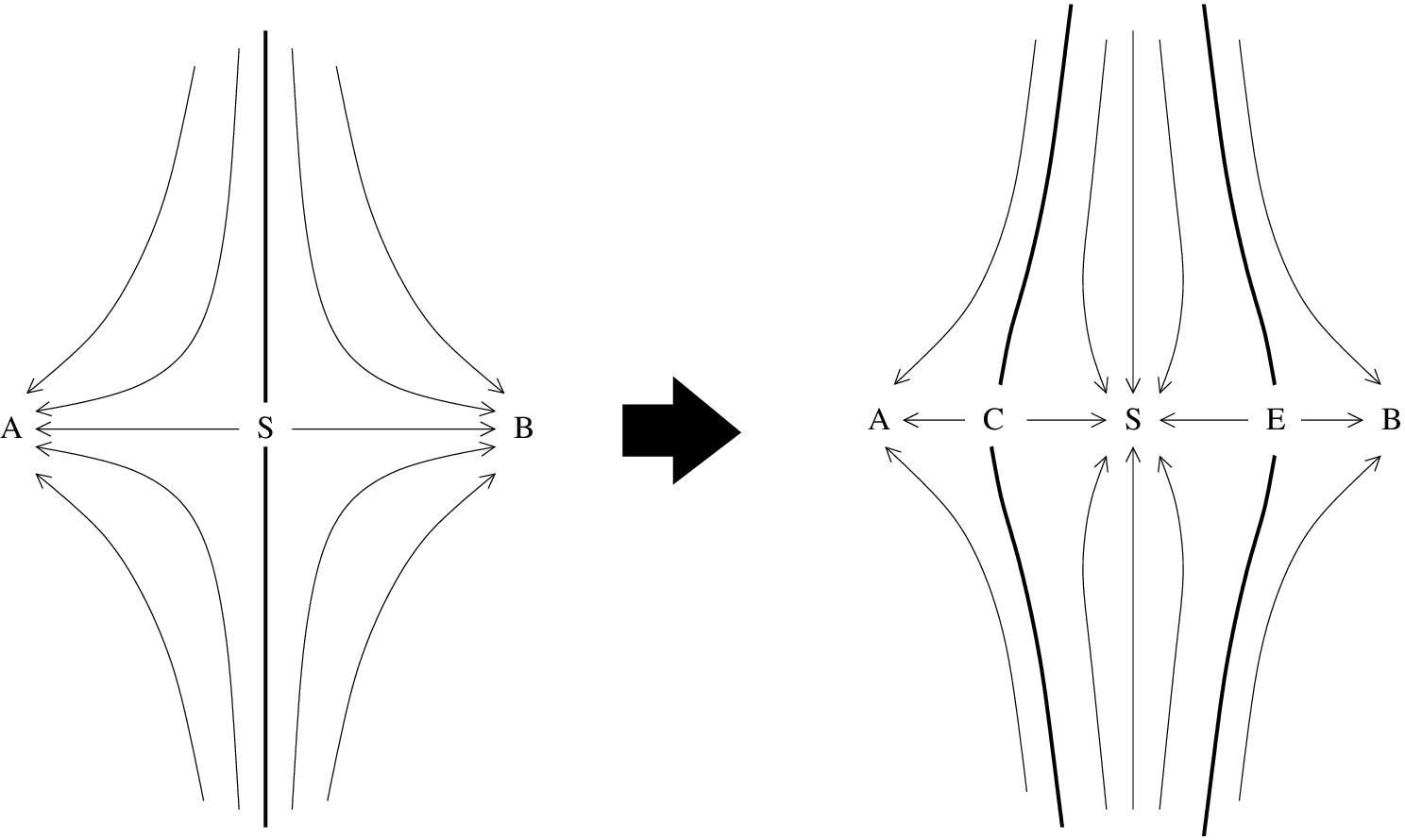}\hfil}
\caption{ 
Effect on the landscape from switching from the potential $U$ to the
pseudo-potential $\vert \vec \nabla U \vert^2$: the saddle $S$ becomes
a minimum with a finite basin of attraction, the basins of attraction
of the minima $A$ and $B$ become smaller. Note that the picture is a
two-dimensional sketch, so that the border of the basins ``acquires''
a small volume: in higher dimensions the effect is enanced.}
\label{figSADDLE}
\end{figure}

\newpage

\begin{figure}
\hbox to\hsize{\epsfxsize=0.8\hsize\hfil
\epsfbox{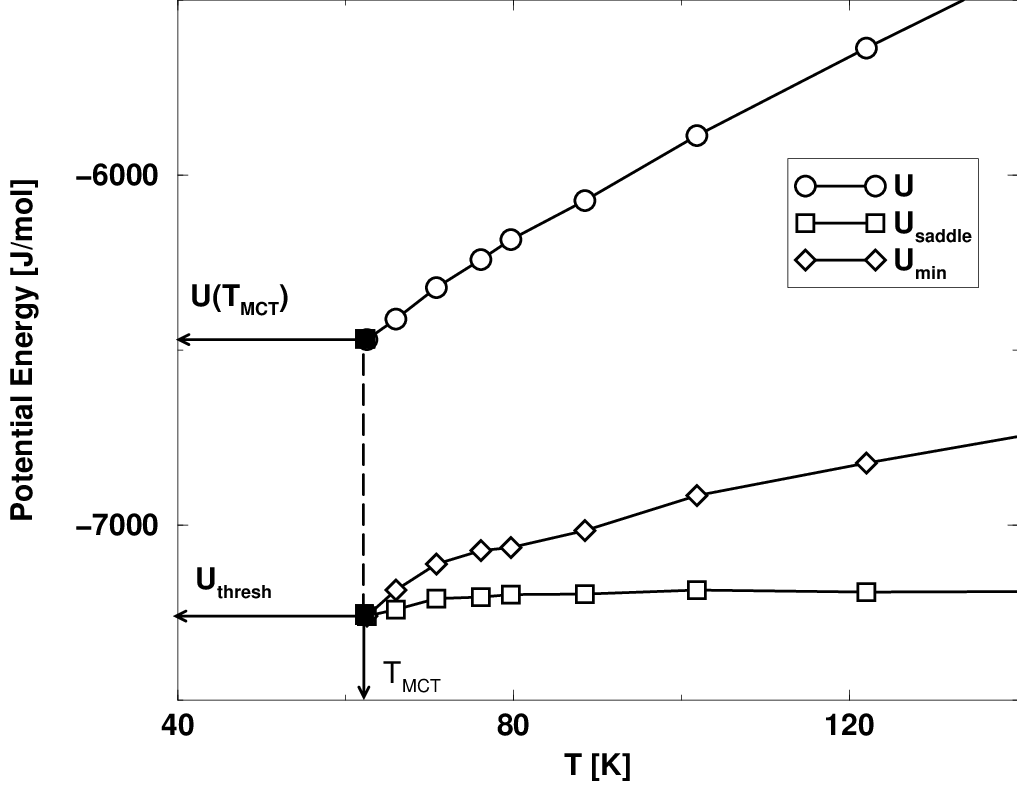}\hfil}
\hbox to\hsize{\epsfxsize=0.4\hsize\hfil
\epsfbox{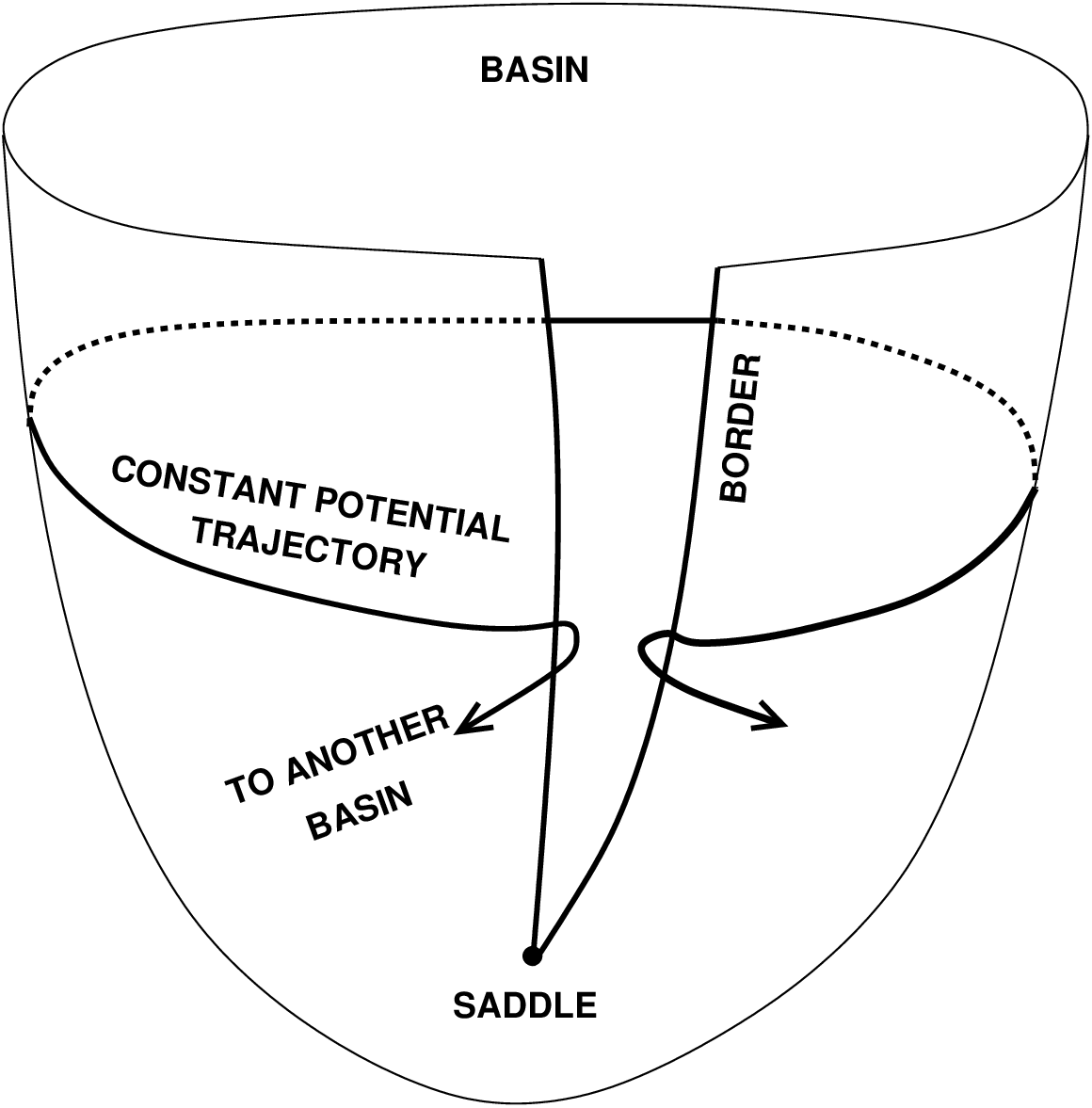}\hfil}
\caption{ 
a) Temperature dependence of the potential energy of the system, of
the saddles and of the minima for the Modified Lennard-Jones system of
ref.~\protect\cite{angesaddle}. The full squares indicate
extrapolations of the energies at $T=T_{MCT}$. Note that, in the range
of the graph, $U(T_{MCT})$ is still much above the potential energy
range of the saddles. b)~Schematic representation of the motion of the
system at a temperature slightly below $T_{MCT}$. For a big system,
the motion happens essentially along constant potential
energy. Although there are still saddles below the trajectory, the
system is seldom near the border above the saddles; when the
trajectory meets the border, a change of basin happens without energy
activation.  }
\label{figBASIN}
\end{figure}
\newpage

\newpage


\begin{references}
\bibitem{funcformINM}
In most of the system studied, the diffusion coefficient $D$ followd
either the law $D$~$\propto$~$<$$f_u$$>$ or the law $D \propto
<\omega_u f_u>$, where $\omega_u$ is the average frequency of
imaginary modes corresponding to the negative eigenvalues of the
Hessian matrix. See (put list of INM).
\bibitem{nota_A}
Obviously the concept of "near" to the border should be rigorously
defined. As shown in ref.~\protect\cite{kurchanlaloux}, in the
thermodynamic limit $N \rightarrow infty$ most of the points are
within a distance $\sim 1/N$ to the border and therefore ``as near as
we wish''.
\bibitem{statpoint}
We remind the reader that the stationary points in question are the
points where the potential energy has a zero derivative $\vec \nabla U
= \vec 0$. These points are characterized by an order $i$ equal to the
number of negative eigenvalues of the Hessian matrix $\nabla^2 U$
(number of directions with negative curvatures). These points are also
called saddles of orther $i$, with the proviso that the saddles of
order $0$ are the minima and the saddles of order $N$ are the maxima.
\end{references}
\end{document}